\documentclass[prb,twocolumn,floatfix,amsmath,amssymb,showpacs]{revtex4}

\usepackage{graphicx}
\usepackage{dcolumn}
\usepackage{bm}

\begin{document}

\title{Superconductivity and local-moment magnetism in
Eu(Fe$_{0.89}$Co$_{0.11}$)$_{2}$As$_{2}$}

\author{Shuai Jiang, Hui Xing, Guofang Xuan, Zhi Ren, Cao Wang, Zhu-an Xu}

\author{Guanghan Cao}
\altaffiliation{Corresponding author} \email{ghcao@zju.edu.cn}
\affiliation{Department of Physics, Zhejiang University, Hangzhou
310027, China\\}

\date{\today}

\begin{abstract}
We report the measurements of resistivity and magnetization under
magnetic fields parallel and perpendicular to the basal plane,
respectively, on a cobalt-doped
Eu(Fe$_{0.89}$Co$_{0.11}$)$_{2}$As$_{2}$ single crystal. We observed
a resistivity drop at $T_c\sim$ 21 K, which shifts toward lower
temperatures under external fields, suggesting a superconducting
transition. The upper critical fields near $T_c$ show large
anisotropy, in contrast with those of other '122' FeAs-based
superconductors. Low-field magnetic susceptibility data also show
evidence of superconductivity below 21 K. Instead of expected
zero-resistance below $T_c$, however, a resistivity reentrance
appears at 17 K under zero field, coincident with the magnetic
ordering of Eu$^{2+}$ moments. Based on the temperature and field
dependences of anisotropic magnetization, a helical magnetic
structure for the Eu$^{2+}$ spins is proposed. External magnetic
fields easily changes the helimagnetism into a ferromagnetism with
fully polarized Eu$^{2+}$ spins, accompanying by disappearance of
the resistivity reentrance. Therefore, superconductivity coexists
with ferromagnetic state of Eu$^{2+}$ spins under relatively low
magnetic field. The magnetic and superconducting phase diagrams are
finally summarized for both $H\parallel ab$ and $H\parallel c$.
\end{abstract}


\maketitle

\section{\label{sec:level1}Introduction}

Superconductivity (SC) and ferromagnetism (FM) are mutually
antagonistic cooperative phenomena, because superconducting state
expels magnetic flux (Meissner effect) but FM generates the internal
magnetic field. On one hand, the internal field generated by FM
destroys SC in two ways: orbital effect\cite{ginzberg57} and
paramagnetic effect (in the case of spin-singlet SC)\cite{paramag}.
On the other hand, SC does not favor FM since SC state suppresses
the zero wave-vector component of the electronic susceptibility,
$\chi(0)$, which is crucial to mediate the localized moments via the
RKKY interaction. The incompatible nature of SC and local-moment FM
was demonstrated in ErRh$_4$B$_4$\cite{fertig77} and
Ho$_{1.2}$Mo$_6$S$_8$\cite{ishikawa} which show destruction of SC at
the onset of long-range magnetic order. Later the repulsive effects
between SC and FM were observed in a family of layered compounds
$R$Ni$_2$B$_2$C ($R$=Tm, Er, Ho and Dy)\cite{eisaki}. The interplay
of SC and FM was also reported in Ru-layer-containing cuprates,
where magnetic ordering temperatures are much higher than SC
transition temperatures.\cite{felner,bernhard} Interestingly, SC and
local-moment\cite{note} FM could be reconciled by considering their
difference in interaction length scale. Earlier theoretical
work\cite{anderson59} pointed out that SC could coexist with
modulated FM such as spiral/helical magnetic configuration or
multidomain structure. Later, it was theoretically shown that SC
could be in the form of spontaneous vortex
state\cite{vortex1,vortex2} to facilitate the FM ordering. However,
there have been few experimental evidences on how SC coexists with
the FM.\cite{review}

Doped EuFe$_2$As$_2$ system is another candidate for searching the
coexistence of SC and local-moment FM. This material consists of two
subsystems: (1) anti-fluorite-type Fe$_{2}$As$_{2}$ layers
responsible for occurrence of superconductivity, and (2)
local-moment-carrying Eu$^{2+}$ ions sandwiched by the
Fe$_{2}$As$_{2}$ layers. In the undoped parent compound
EuFe$_2$As$_2$, the two subsystems undergoes an antiferromagnetic
(AFM) spin-density wave (SDW) transition associated with Fe moments
at 190 K and another AFM ordering for Eu$^{2+}$ spins at 19 K,
respectively.\cite{Eu122,Ren,Jeevan1,Wu} The magnetic structure of
the latter AFM order was proposed to be of A-type,\cite{Jiang} in
which Eu$^{2+}$ spins algin ferromagnetically in the basal planes
but antiferromagnetically along the $c$-axis, based on the
anisotropic magnetic and magnetotransport measurements. This
magnetic structure was very recently confirmed by the magnetic
resonant x-ray scattering (Ref.~\cite{XAS}) and neutron diffraction
(Ref.~\cite{ND}) experiments.

By the partial substitution of Eu with K, SC over 30 K was reported
in Eu$_{1-x}$K$_{x}$Fe$_2$As$_2$.\cite{Jeevan2} However, no magnetic
ordering for Eu$^{2+}$ spins was observed, probably due to the
dilution effect by the Eu-site doping. In the case of Fe-site
doping, though superconductivity at 20 K was obtained in
BaFe$_{2-x}$Ni$_x$As$_2$ (Ref.~\cite{122Ni}), attempt to obtain SC
in EuFe$_{2-x}$Ni$_x$As$_2$ was unsuccessful.\cite{Ren-Ni} Instead,
the Ni doping leads to FM ordering for the Eu$^{2+}$ moments. By
phosphorus doping at the As-site, which also keeps Eu$^{2+}$
sublattice undisturbed, we found bulk SC at $T_c$=26 K followed by a
local-moment FM at 20 K in
EuFe$_{2}$(As$_{0.7}$P$_{0.3}$)$_{2}$.\cite{Ren-P} In fact, with
applying pressure, superconductivity at 29 K was reported in the
undoped EuFe$_2$As$_2$,\cite{Miclea,Terashima} where the AFM
ordering for Eu$^{2+}$ moments was proposed. The above results
suggest that the prerequisite for finding the coexistence of SC and
local-moment magnetism in Eu-containing arsenides is that $T_c$
should be higher than the magnetic ordering temperature $T_{M}$.
Note that the maximum $T_c$ in BaFe$_{2-x}$Co$_x$As$_2$ is as high
as 25 K,\cite{122Co} therefore, we investigated the
Eu(Fe$_{1-x}$Co$_x$)$_2$As$_2$ system. Consequently, evidence of SC
transition was observed for 0.09$\leq x <$0.15, basically consistent
with a very recent report by Zheng et al.\cite{CXH}

In this paper, we present detailed measurements of the resistivity
and magnetization under magnetic fields using well-characterized
single crystals of Eu(Fe$_{0.89}$Co$_{0.11}$)$_{2}$As$_{2}$. We
observed a resistivity drop at 21 K for both in-plane resistivity
($\rho_{ab}$) and out-plane resistivity ($\rho_{c}$), which is
ascribed to a SC transition. Evidence of superconductivity is also
given by low-field magnetic susceptibility measurement. Followed by
the SC transition, a resistivity reentrance appears as the Eu$^{2+}$
spins order spontaneously. By analyzing the temperature and field
dependences of anisotropic magnetization, and comparing with the
magnetic structure of EuFe$_2$As$_2$, a helical magnetic structure
for Eu$^{2+}$ spins was proposed. External magnetic field
re-orientates the Eu$^{2+}$ moments easily, changing the
helimagnetism into ferromagnetism. Finally, the magnetic and
superconducting phase diagrams were established, exhibiting the
intriguing coexistence of SC and long-range magnetic ordering in
Eu(Fe$_{0.89}$Co$_{0.11}$)$_{2}$As$_{2}$.

\section{\label{sec:level2}Experimental}

Single crystals of Eu(Fe$_{1-x}$Co$_{x}$)$_{2}$As$_{2}$ were grown
using (Fe,Co)As as the self flux, similar to previous
reports\cite{CXH-SX,Jiang}. (Fe,Co)As with the atomic ratio
Fe:Co=$(1-x):x$ was presynthesized by reacting Fe powders with As
shots in vacuum at 773 K for 6 h and then at 1030 K for 12 h. Fresh
Eu grains and Fe$_{1-x}$Co$_{x}$As powders were thoroughly mixed in
a molar ratio of 1:4. The mixture was loaded into an alumina tube,
then put into a quartz ampoule. The sealed quartz ampoule was heated
to 1053 K at a heating rate of 150 K/h holding at this temperature
for 10 h. Subsequently, the temperature was raised to 1398 K in 3 h
holding for 5 h. The crystals were grown by slowly cooling to 1223 K
at a cooling rate of 2 K/h. Finally, the quartz ampoule was cooled
to room temperature by shutting off the furnace. Many shiny
plate-like crystals with the typical size of $3\times2\times0.1$
mm$^3$ were obtained.

The crystals were characterized by x-ray diffraction (XRD) and
field-emission scanning electron microscopy (SEM), and energy
dispersive x-ray (EDX) spectroscopy. XRD was performed using a
D/Max-rA diffractometer with Cu-K$\alpha$ radiation and a graphite
monochromator. SEM image was taken in a field-emission scanning
electron microscope (Sirion FEI, Netherlands) equipped with a
Phoenix EDAX x-ray spectrometer. Figure 1 shows the morphological,
compositional and structural characterizations on a Co-doped
EuFe$_2$As$_2$ crystal. The SEM image of the crystal measured shows
large area of flat surfaces with only minor impurities adhered to.
Quantitative analysis for the EDX spectra indicates that the
composition is Eu(Fe$_{0.89}$Co$_{0.11}$)$_{2}$As$_{2}$ within the
measurement error ($\pm5\%$). XRD pattern of $\theta-2\theta$ scan
shows only (00l) reflections, indicating that the $c$-axis is
perpendicular to the crystal sheet planes. The $c$-axis was
calculated to be 1.207 nm which is reasonably smaller than that of
EuFe$_2$As$_2$ (Ref.~\cite{Ren}). The rocking curve ($\theta$ scan)
shown in the inset has a relatively small Full Width at Half Maximum
(FWHM), suggesting high quality of the sample.

\begin{figure}
\includegraphics[width=8cm]{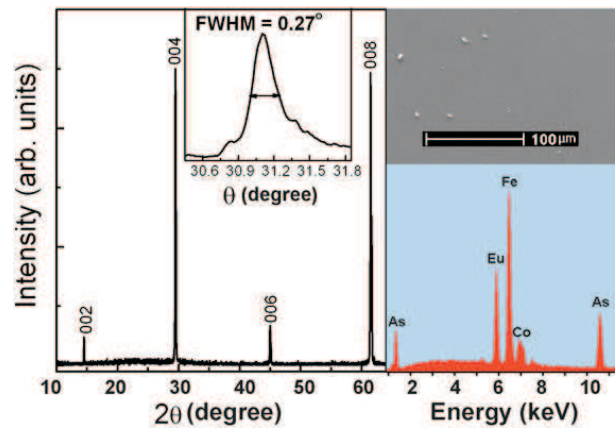}
\caption{(Color online) Characterizations of a Co-doped
EuFe$_2$As$_2$ crystal in the present study by (a) x-ray
diffraction, (b) scanning electron microscope and (c) energy
dispersive x-ray spectroscopy.}
\end{figure}

Electrical resistivity was measured using a standard four-terminal
method. The electrode configuration in Ref.~\cite{CXH-SX} was
employed for measuring $\rho_c$. The dc magnetization was measured
on a Quantum Design magnetic property measurement system (MPMS-5).
The crystal was carefully mounted on a sample holder, with the
applied field perpendicular or parallel to the crystallographic
$c$-axis. The deviation angle was estimated to be less than
5$^\circ$.

We found that the SDW transition in the parent compound was
suppressed with the Co doping, like the cases in other iron
arsenides.\cite{122Co,Co} For 0.09$\leq x <$0.15, resistivity drop
due to a SC transition was observed around 20 K. The sample of
$x$=0.09 showed a resistivity upturn at 30 K due to the residual SDW
transition. For the sample with $x$=0.11, no clear evidence of SDW
transition could be observed. Compared with the
Ba(Fe$_{1-x}$Co$_{x}$)$_{2}$As$_{2}$ system,\cite{122Co} the optimal
doping level in Eu(Fe$_{1-x}$Co$_{x}$)$_{2}$As$_{2}$ shifts to a
larger value. In this paper we focus on the physical property
measurements for the optimally doped sample with $x$=0.11.

\section{\label{sec:level3}Results and discussion}

\subsection{Resistivity}

Figure 2 shows $\rho_{ab}$ and $\rho_{c}$ for
Eu(Fe$_{0.89}$Co$_{0.11}$)$_{2}$As$_{2}$ crystals under zero field.
While $\rho_{c}$ is nearly 50 times large of $\rho_{ab}$, their
temperature dependences are almost the same. At high temperatures
both show usual metallic behavior. Around 20 K the resistivity drops
by over 30\%, suggesting a SC transition. However, it increases
sharply below $T_{\text{ret}}$ = 17 K ($T_{\text{ret}}$ denotes
resistivity reentrance temperature), and a resistivity peak appears
at 16 K. One notes that the resistivity maximum is still much lower
than that of the undoped EuFe$_2$As$_2$, as shown in the upper inset
of Fig. 2. This implies that the state around 16 K is still within
the SC regime. At lower temperatures, the resistivity tends to
saturate at a residual value. This result resembles the behavior of
EuFe$_2$As$_2$ under high pressures,\cite{Miclea} which was ascribed
as a reentrant superconductivity. The two transitions can also be
manifested by the anomalous peaks in $\rho_{c}/\rho_{ab}$, shown in
the lower inset of Fig. 2.

\begin{figure}
\includegraphics[width=8cm]{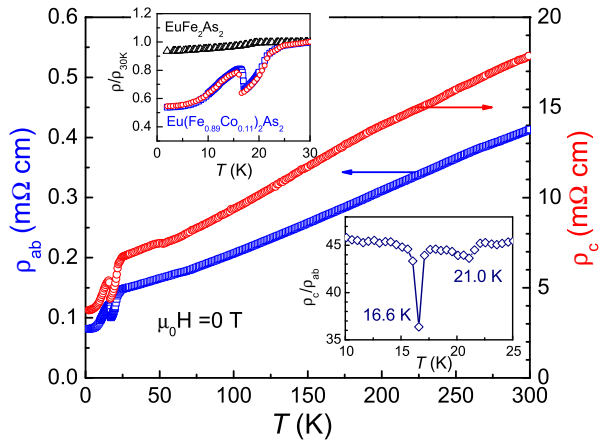}
\caption{(Color online) Temperature dependence of in-plane and
out-plane resistivity for Eu(Fe$_{0.89}$Co$_{0.11}$)$_{2}$As$_{2}$
crystals at zero field. Upper left inset is an expanded plot in
comparison with the data of the nonsuperconducting
EuFe$_{2}$As$_{2}$crystals. Lower right inset displays the
anisotropic ratio $\rho_{c}/\rho_{ab}$, showing two peaks associated
with superconducting and magnetic transitions, respectively.}
\end{figure}

To clarify the above two resistivity anomalies, we performed the
magnetoresistance measurements. Fig. 3(a) shows the in-plane
resistivity under magnetic fields parallel to the basal planes
(hereafter denoted by $H\parallel ab$). As expected for a SC
transition, the resistivity drop shifts to lower temperatures with
increasing magnetic fields. On the other hand, the resistivity peak
is drastically suppressed by the applied fields. When the applied
field is perpendicular to the basal planes, as shown in Fig. 3(b),
the SC transition is suppressed more severely by the field. However,
the resistivity peak is not influenced very much until it is
'buried' by the SC transition. The inset of Fig. 4 clearly shows the
different response of the $T_{\text{ret}}$ to the applied field
along different directions. This observation is in sharp contrast
with that in \emph{R}Ni$_{2}$B$_{2}$C$_{2}$
superconductors,\cite{eisaki} where the reentrant region becomes
much enlarged by the external field.

\begin{figure}
\includegraphics[width=8cm]{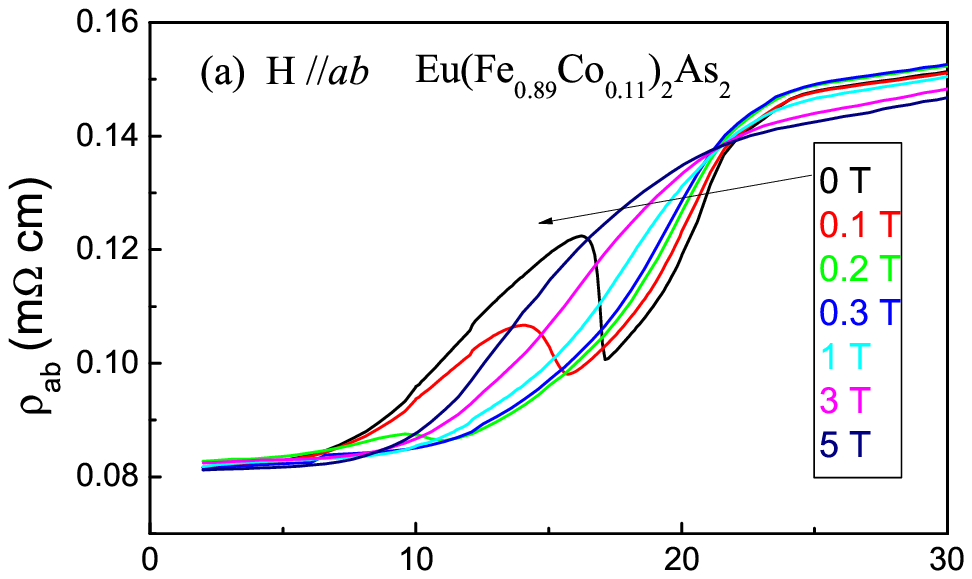}
\includegraphics[width=8cm]{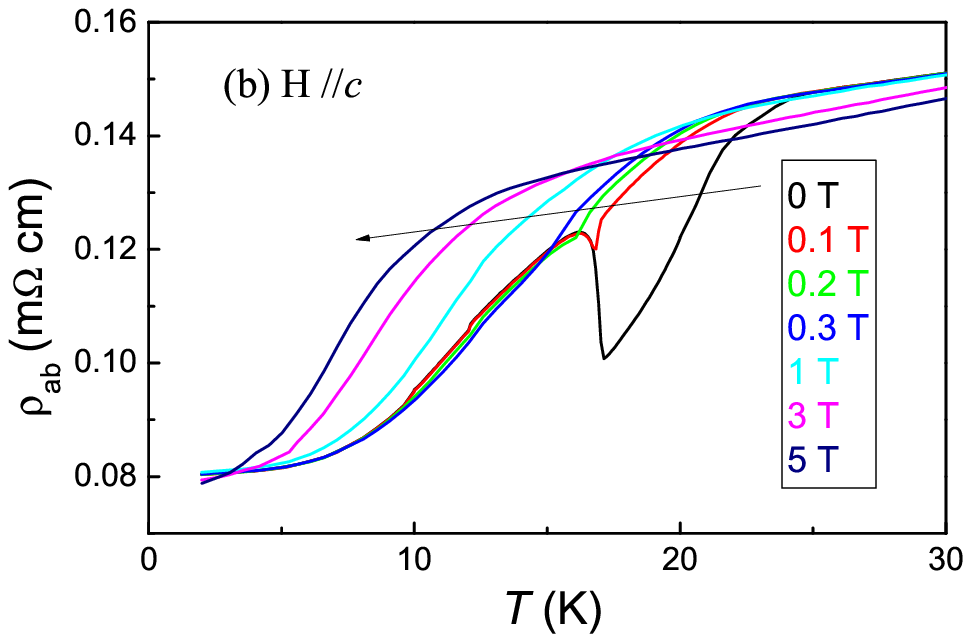}
\caption{(Color online) Resistive transition under magnetic fields
for Eu(Fe$_{0.89}$Co$_{0.11}$)$_{2}$As$_{2}$ crystals.
(a)$H\parallel ab$; (b)$H\parallel c$.}
\end{figure}

From the magnetoresistivity data, the upper critical fields were
determined by using the criterion of 90\% normal-state resistivity.
As shown in Fig. 4, upward curvature can be seen in the $H_{c2}(T)$
curves, especially for $H\perp ab$. The anisotropic ratio,
$H_{c2}^{\parallel}$/$H_{c2}^{\perp}$, achieves 30 at $\sim$ 17 K.
This contrasts with the nearly isotropic $H_{c2}$ in
BaFe$_2$As$_2$.\cite{Yuan} The large anisotropy in $H_{c2}$ reflects
the interplay between SC and magnetic ordering of Eu$^{2+}$ moments.
The initial slope
$\mu_{0}$$\partial$$H_{c2}^{\parallel}$/$\partial$$T$ near $T_{c}$
is $-1.3$ T/K, giving an upper critical field of
$\mu_{0}H_{\text{c2}}^{\parallel}$(0) $\sim$ 26 T by linear
extrapolation. This upper critical field is obviously lower than the
Pauli paramagnetic limit $\mu_{0}H_{P}=1.84 T_c\approx$ 38.6 T. The
situation is similar to that in the
EuFe$_{2}$(As$_{0.7}$P$_{0.3}$)$_{2}$ superconductor
(Ref.~\cite{Ren-P}), but different from those of other Eu-free
ferroarsenide superconductors (Ref.~\cite{exceed-Pauli}). The lower
magnitude of $H_{c2}(0)$ specially in Eu-containing superconductors
implies the existence of significant internal field from the
Eu$^{2+}$ moments.

\begin{figure}
\includegraphics[width=8cm]{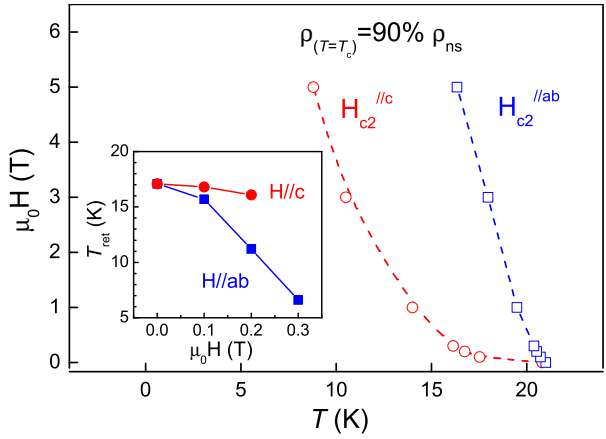}
\caption{(Color online) Upper critical fields of
Eu(Fe$_{0.89}$Co$_{0.11}$)$_{2}$As$_{2}$ single crystals. Inset:
resistivity reentrance temperature $T_{\text{ret}}$ as a function of
applied field.}
\end{figure}

Figure 5 shows the isothermal resistivity under magnetic field
parallel or perpendicular to the basal planes. At 30 K, the
resistivity decreases monotonically with the field. Negative
magnetoresistance (MR) was also observed in EuFe$_2$As$_2$ just
above the Eu-AFM ordering temperature,\cite{Jiang} which was
ascribed to the reduction of Eu-spin disorder scattering by the
external magnetic field. At 21 and 17 K, an abrupt increase in
resistivity at relatively low fields, especially for $H\perp ab$,
representing the transition from superconductivity to normal state.
The normal-state $\rho_{ab}^{\parallel}$ increases with the field,
which reflects the intrinsic transport property of FeAs layers,
because of field-induced ferromagnetic transition. At 10 K, SC
coexists with the helical magnetic order (see the next section) at
low fields. The resistivity first decreases to a minimum at $H^*$
then increases again with the field. The decrease in $\rho_{ab}$ is
related to the reorientation of Eu$^{2+}$ moments, because $H^*=H_s$
($H_s$ refers to the saturated field, see Fig. 9 in the next
section). The increase in $\rho_{ab}$ is probably due to the
increase of SC vortices by the external field and/or the intrinsic
transport property of FeAs layers. At 2 K and 4 K,
$\rho_{ab}^{\perp}$ first increases to a maximum at $H^*$ and then
starts to decrease with the field. In the case of $H\parallel ab$,
$\rho_{ab}^{\parallel}$ first increases also, then decrease to a
minimum at $H^*=H_s$. Interestingly, another maximum appears at
higher field. These data should reflect the interplay between SC and
magnetism, but we fail to have a sound explanation at present. The
non-zero resistance is probably due to the dissipation of the motion
of spontaneous vortex, generated by the magnetic ordering of
Eu$^{2+}$ spins. However, such spontaneous vortex should be directly
evidenced before a quantitative understanding.

\begin{figure}
\includegraphics[width=8cm]{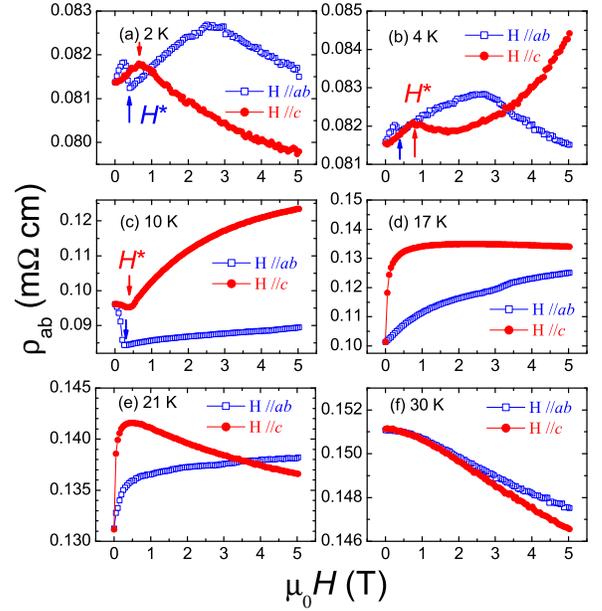}
\caption{(Color online) Field dependence of in-plane resistivity in
Eu(Fe$_{0.89}$Co$_{0.11}$)$_{2}$As$_{2}$ at fixed temperatures. The
turning in resistivity at $H^*$ are marked by arrows.}
\end{figure}

\subsection{Magnetic Properties}

Figure 6 shows the temperature dependence of magnetic
susceptibility. The high temperature susceptibility well obeys
Curie-Weiss behavior: $\chi=\chi_{0}+C/(T-\theta)$, where $\chi_0$
denotes the temperature-independent term, $C$ the Curie-Weiss
constant and $\theta$ the paramagnetic Curie temperature. The data
fitting (50 K $<T<$ 200 K) shows that the effective moment is close
to the theoretical value $g\sqrt{S(S+1)}\mu_{B}$=7.94 $\mu_{B}$
($S=7/2$ and $g$=2) for a free Eu$^{2+}$ ion. The $\theta$ values
are positive, suggesting ferromagnetic interaction among Eu$^{2+}$
spins.

\begin{figure}
\includegraphics[width=8cm]{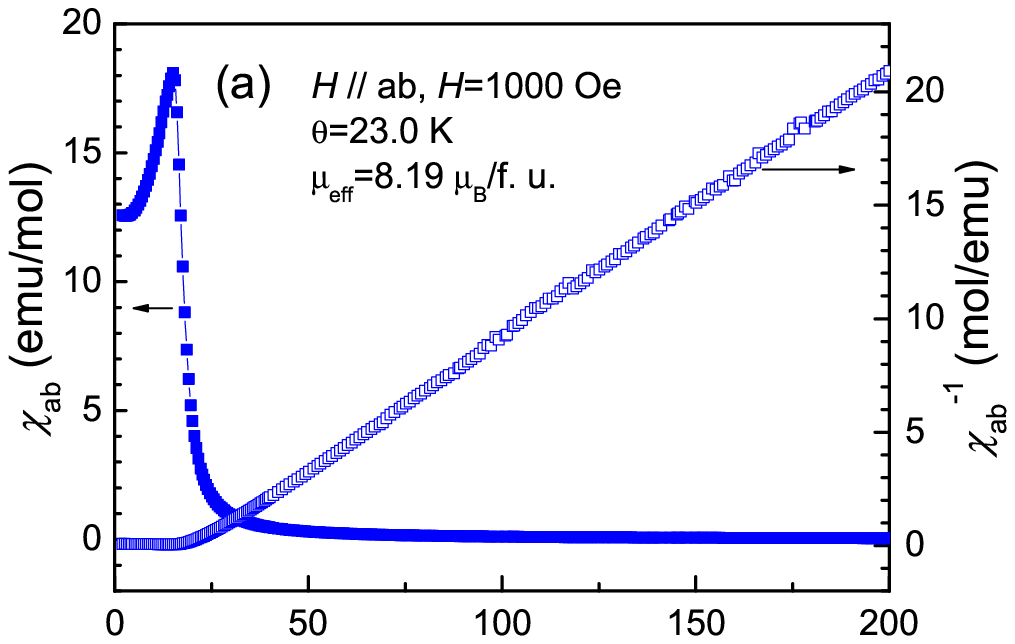}
\includegraphics[width=8cm]{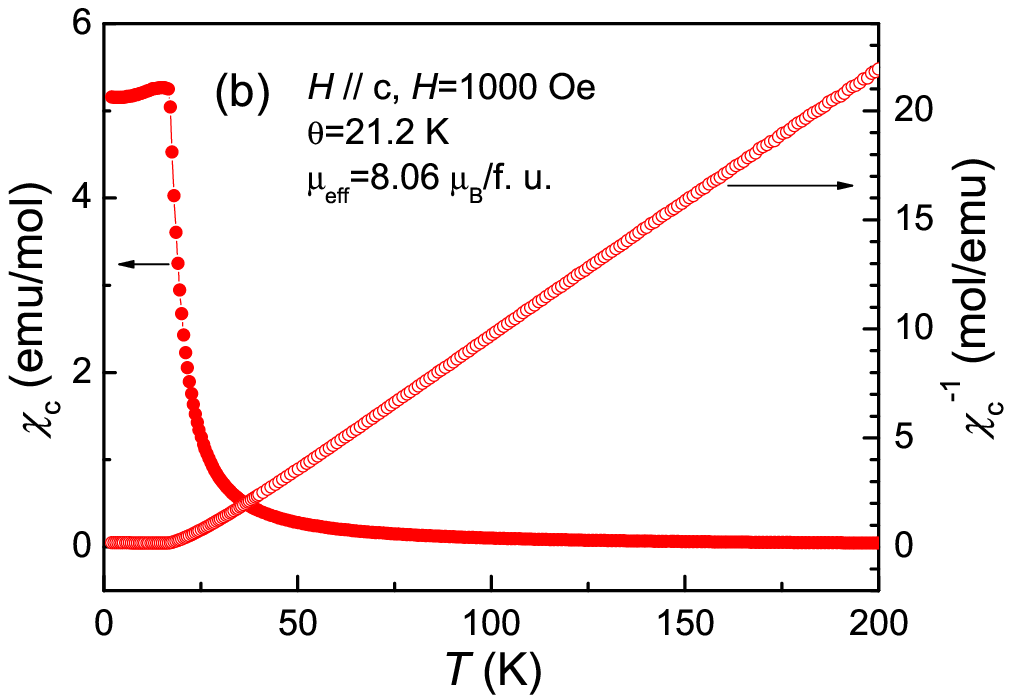}
\caption{(Color online) Temperature dependence of magnetic
susceptibility for Eu(Fe$_{0.89}$Co$_{0.11}$)$_{2}$As$_{2}$. The
measurements were performed in field cooling mode under applied
field of 1000 Oe. (a)$H\parallel ab$; (b)$H\parallel c$.}
\end{figure}

Though the high-temperature susceptibility is basically isotropic,
$\chi_{ab}$ is obviously higher than $\chi_{c}$ at low temperatures,
e.g., $\chi_{ab}/\chi_{c}$ is about 3.5 at 17 K. This suggests that
the easy magnetization direction is parallel to the $ab$ planes,
similar to the case in EuFe$_2$As$_2$.\cite{Jiang} Below $T_M$ = 17
K, $\chi_{ab}$ decreases rather sharply, indicating an
antiferromagnetic-like transition. On the other hand, $\chi_{c}$
remains nearly constant below $T_M$. Therefore, one concludes that
the Eu$^{2+}$ moments are perpendicular to the $c$-axis below $T_M$.
Considering the dominate ferromagnetic interaction among Eu$^{2+}$
spins, one expects ferromagnetic arrangement for the Eu$^{2+}$ spins
within single Eu$^{2+}$ layer. This is quite similar to the
situation in EuFe$_2$As$_2$ (Ref.~\cite{Jiang}), in latter case the
magnetic ordering temperature is 2 K higher.

However, we note that the magnitude of drop in $\chi_{ab}$ is much
smaller, compared with EuFe$_2$As$_2$ crystals.\cite{Jiang} The
residual susceptibility at zero temperature is about 2/3 of
$\chi_{\text{max}}$ at $T_M$, irrespective of changing the relative
orientation between the sample and the applied field within $ab$
planes. In addition, the field dependence of magnetization shows
only a spin re-orientation process for $H\parallel ab$ (see Fig. 9),
in contrast with the step-like magnetization curves in
EuFe$_2$As$_2$ (Ref.~\cite{Jiang}). Both results suggest the
non-collinear alignment for Eu$^{2+}$ spins, though lying in the
$ab$ planes. Therefore, we propose a helical magnetic order for
Eu$^{2+}$ moments in Eu(Fe$_{0.89}$Co$_{0.11}$)$_{2}$As$_{2}$, i.e.,
the moments of the neighboring FM Eu$^{2+}$ layers form an angle of
$\varphi$ ($\varphi \neq n\pi$, $n$ is an integer). Such a
non-collinear magnetic order was first observed in 1950s in
MnAu$_2$,\cite{MnAu2} in which the FM basal planes of Mn atoms are
sandwiched by two layers of Au atoms.

The Eu-interlayer spacing is so large that interlayer magnetic
coupling should be an indirect RKKY interaction, which has much
longer range and changes its sign with the distance and Fermi wave
vector. In the frame work of RKKY interaction, the above
non-collinear helimagnetism (HM) is possible if considering both
nearest neighboring (NN) and next nearest neighboring (NNN) (along
the $c$-axis) interlayer couplings. According to a simplified
derivation,\cite{theta}
\begin{equation}
\text{cos}\varphi=-\frac{J_{\text{NN}}}{4J_{\text{NNN}}}.\label{theta}
\end{equation}
The above solution corresponds to helimagnetic order, when
$|J_{\text{NN}}|<|4J_{\text{NNN}}|$. Here we note that the HM is
compatible with SC order, as theoretical work\cite{anderson59}
pointed out.

\begin{figure}
\includegraphics[width=8cm]{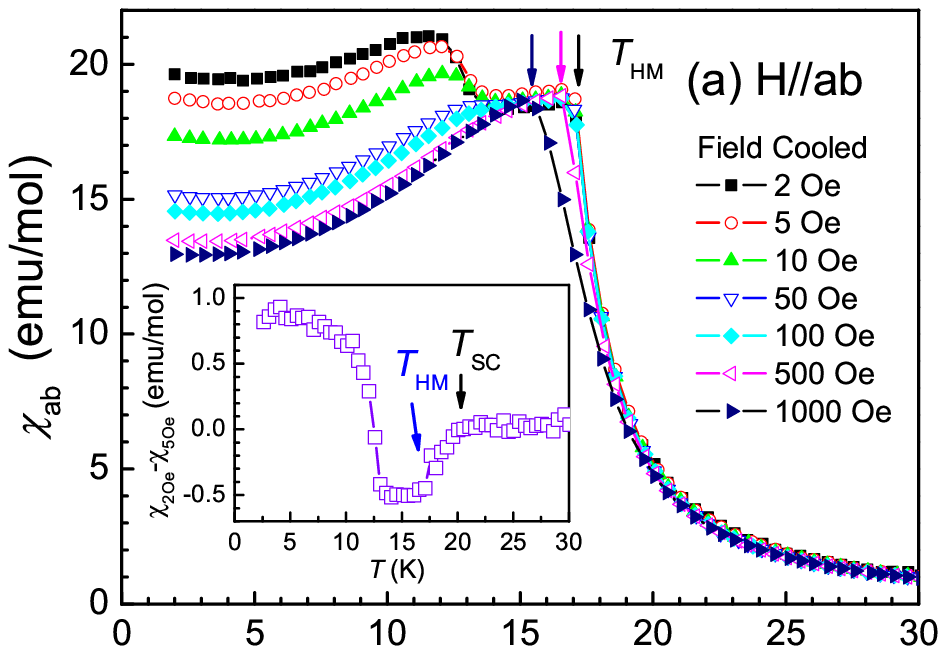}
\includegraphics[width=8cm]{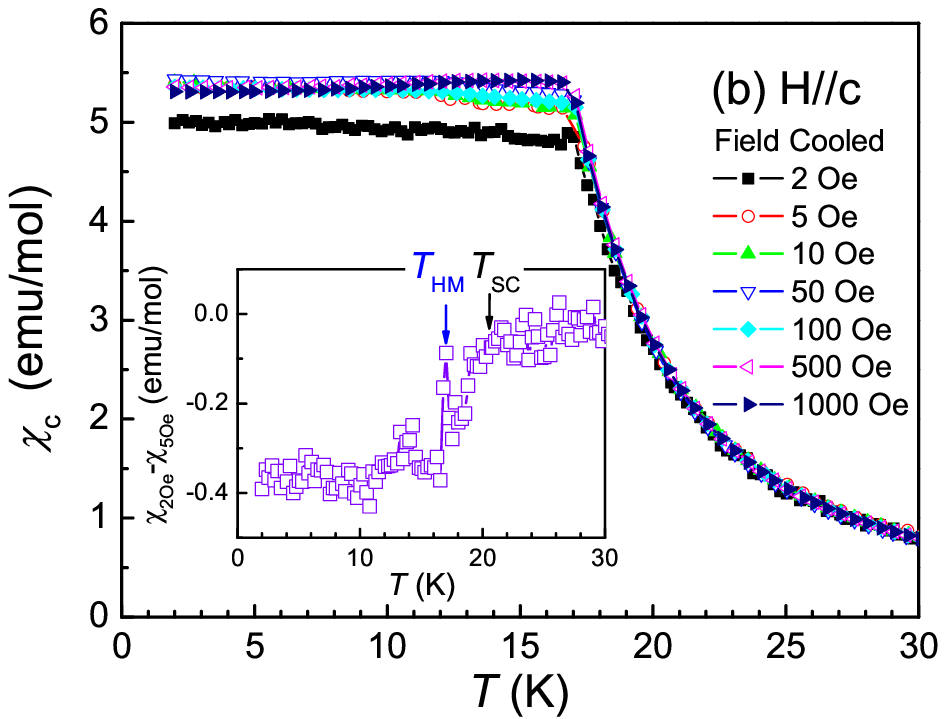}
\caption{(Color online) Low-field magnetic susceptibility for
Eu(Fe$_{0.89}$Co$_{0.11}$)$_{2}$As$_{2}$. The insets make a
subtraction: $\Delta\chi$=$\chi_{2\text{Oe}}-\chi_{5\text{Oe}}$. The
superconducting temperature $T_{SC}$ and the helical magnetic
ordering temperature $T_{HM}$ are marked. (a)$H\parallel ab$;
(b)$H\parallel c$.}
\end{figure}

Due to the proximity of superconducting transition and magnetic
ordering, the superconducting diamagnetic signal could be very weak.
The huge paramagnetic background from Eu$^{2+}$ spins also makes it
difficult to directly observe the diamagnetism. To find signal of
SC, we carried out the low-field susceptibility measurement, as
shown in Fig. 7. For $H\parallel ab$, the magnetic transition
temperature decreases even by a small field of 500 Oe. When the
field is less than 10 Oe, an increase in $\chi$ can be observed at
13 K. Such an anomaly is pronounced with decreasing field. Thus we
made a subtraction:
$\Delta\chi$=$\chi_{2\text{Oe}}-\chi_{5\text{Oe}}$, as shown in the
inset. One sees an abrupt decrease at 21 K, corresponding to the
resistivity drop in Fig. 2. This result is reproducible for the
subtractions using different $\chi_{\text{H}}(T)$ data. Furthermore,
the subtraction of $\chi_{\text{FC}}(T)$ from $\chi_{\text{ZFC}}(T)$
also gives evidence of SC below 21 K. The "diamagnetism" in the
paramagnetic background suggests SC in
Eu(Fe$_{0.89}$Co$_{0.11}$)$_{2}$As$_{2}$. The absence of bulk
Meissner effect, similar to the case in
EuFe$_{2}$(As$_{0.7}$P$_{0.3}$)$_{2}$ (Ref.~\cite{Ren-P}), should be
associated with the magnetic ordering of Eu$^{2+}$ spins.
Theoretical work\cite{Varma} indicates that, in the limit of large
saturated magnetic moment and magnetic anisotropy, there will be no
Meissner effect. In that case, the effective lower critical field
$H_{c1}$ will be zero and superconductivity ¡°appears¡± only when
vortices are pinned to impurity sites. In fact, the above difference
in $\chi_{\text{H}}(T)$ for $H$= 2 and 5 Oe suggests that the
$H_{c1}$ is really much lower than expected.

Here we have to address another anomaly in $\Delta\chi$, i.e., the
increase at 13 K. This phenomenon is reminiscent of paramagnetic
Meissner effect (PME). Intrinsic PME can be produced from a
spontaneous flux in a SC loop made of Josephson junction with
superconducting phase difference.\cite{PME} In the SC and HM
coexisted state, similar junctions can be possibly formed due to the
proximity effect in SC-FM boundaries.\cite{rmp2005} Therefore,
spontaneous flux could be generated mostly parallel to $ab$-planes.
This could result in the observed PME for $H\parallel ab$. In the
case of $H\perp ab$, the SC transition at 21 K can also be clearly
seen. However, the PME-like transition is not so obvious, consistent
with the spontaneous flux perpendicular to $c$-axis.

\begin{figure}
\includegraphics[width=8cm]{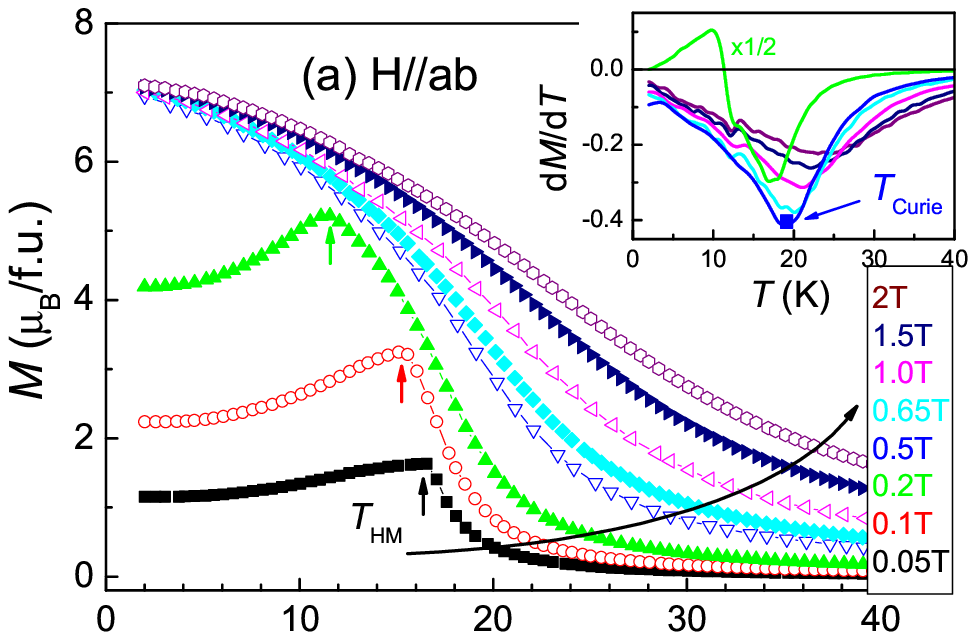}
\includegraphics[width=8cm]{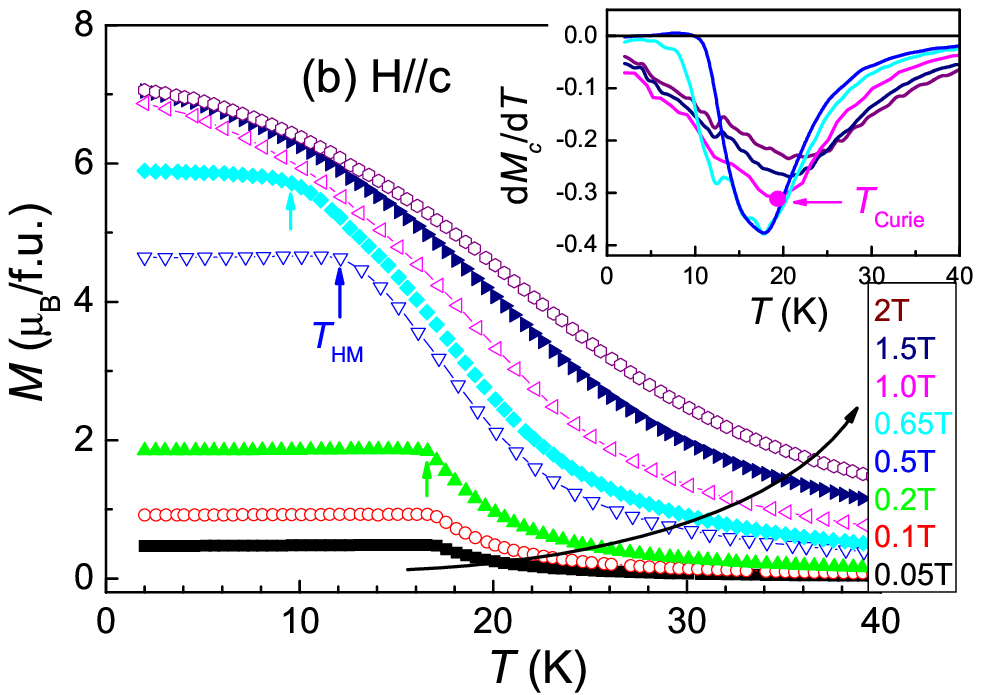}
\caption{(Color online) Temperature dependence of magnetization at
fixed magnetic fields for Eu(Fe$_{0.89}$Co$_{0.11}$)$_{2}$As$_{2}$
crystals. The arrows mark the helimagnetic ordering temperature. The
inset plots the derivative of magnetization, showing the
ferromagnetic transitions at $T_{\text{Curie}}$. (a)$H\parallel ab$;
(b)$H\parallel c$.}
\end{figure}

Figure 8 shows the temperature dependence of magnetization under
fixed magnetic fields. For both $H\parallel ab$ and $H\perp ab$,
$T_{\text{HM}}$ decreases with the field. Compared with
$T_{\text{HM}}^{\perp}$, $T_{\text{HM}}^{\parallel}$ is more easily
suppressed by the magnetic fields. The variations of $T_{\text{HM}}$
coincide with the changes in $T_{\text{ret}}$ (shown in Fig. 3),
suggesting that the resistivity reentrance is closely related to the
helimagnetic transitions. The decrease in $T_{\text{HM}}$ by
external fields can be qualitatively understood in terms of the
above simple model considering the interlayer magnetic couplings
$J_{NN}$ and $J_{NNN}$. Under magnetic fields, the effective
coupling is modified as $J_{\text{eff}}=J+J_{\text{ext}}$
($J_{\text{ext}}$ denotes the contribution from the applied field).
Thus the applied field possibly makes the inequality
$|J_{\text{NN,eff}}|<|4J_{\text{NNN,eff}}|$ invalid (note that
$|J_{\text{NN,eff}}|=|J_{\text{NN}}+J_{\text{ext}}|;
|4J_{\text{NNN,eff}}|=4|J_{\text{NNN}}+J_{\text{ext}}|$), leading to
the appearance of a more stabilized FM phase.

Under higher magnetic field, the HM-FM transition can be verified by
the saturation of magnetization to a fully polarized value $gS$=7.0
$\mu_B$/ f.u. ($g$=2 and $S$=7/2). Here we identify the FM
transition temperature $T_{\text{Curie}}$ as the inflection point of
the $M(T)$ curves. The derivative of magnetization, plotted in the
inset of Fig. 8, indicates that $T_{\text{Curie}}$ increases with
the field.

\begin{figure}
\includegraphics[width=8cm]{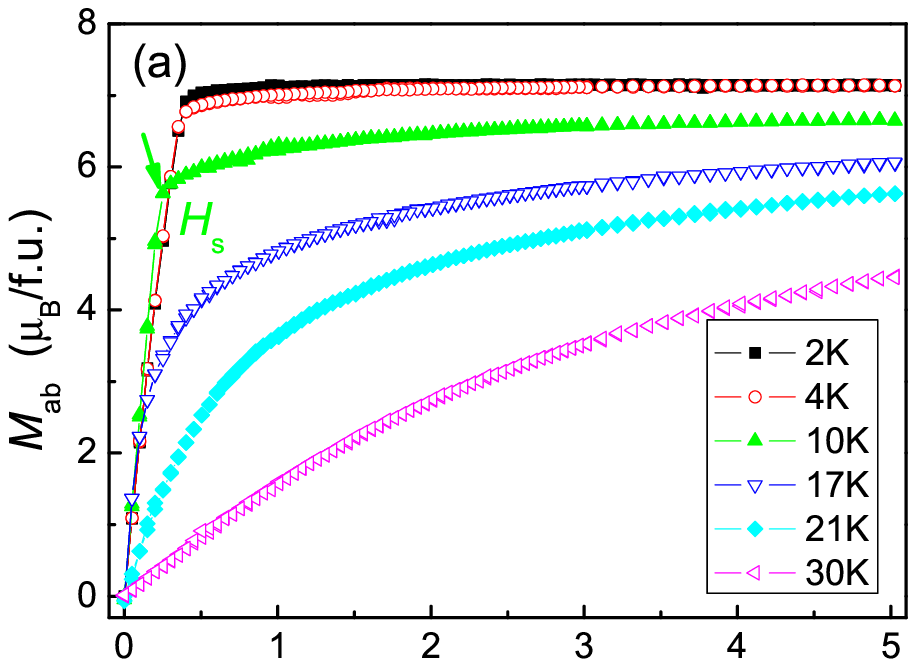}
\includegraphics[width=8cm]{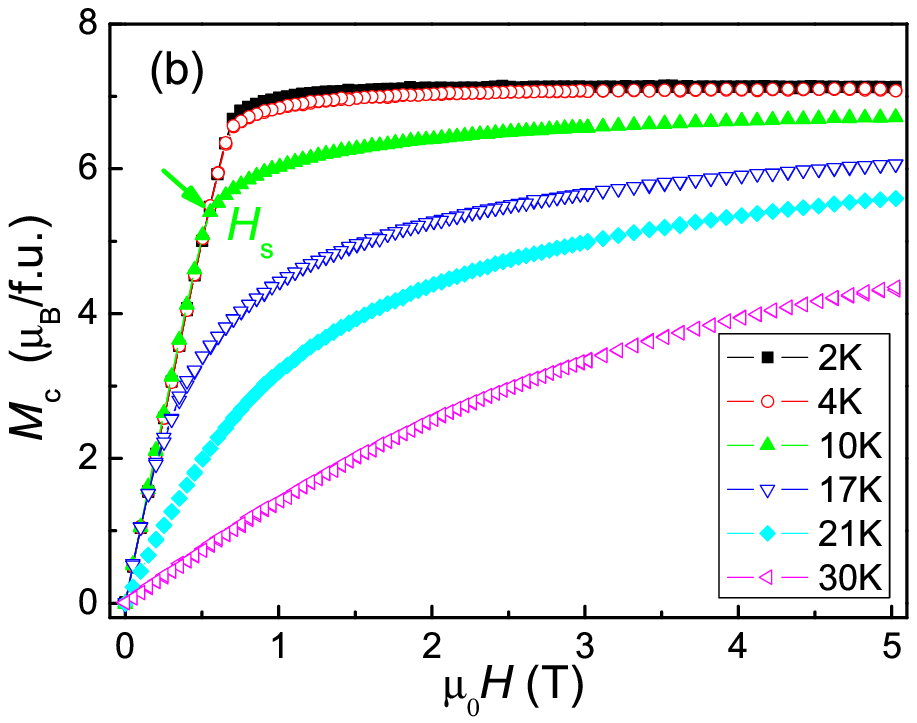}
\caption{(Color online) Field dependence of magnetization at fixed
temperatures for Eu(Fe$_{0.89}$Co$_{0.11}$)$_{2}$As$_{2}$ crystals.
The saturated field $H_{s}$ is marked by the arrow. (a)$H\parallel
ab$; (b)$H\parallel c$.}
\end{figure}

Figure 9 shows the isothermal magnetization for the
Eu(Fe$_{0.89}$Co$_{0.11}$)$_{2}$As$_{2}$ crystals. At 2 K, the
magnetization increases almost linearly until achieving the
saturated value of 7.0 $\mu_B$/ f.u.  for both directions of
magnetic fields. The $M_{c}(H)$ behavior resembles that of
EuFe$_2$As$_2$, except for the smaller saturated field
$H_s^{\perp}$. However, the $M_{ab}(H)$ curve is qualitatively
different from its counterpart of EuFe$_2$As$_2$ crystals. The
latter shows a step-like magnetization at 2 K, which was identified
as a metamagnetic transition associated with a spin-flip
process.\cite{Jiang} Since the spin flip is related to the A-type
antiferromagnetic structure, the absence of step-like magnetization
in Eu(Fe$_{0.89}$Co$_{0.11}$)$_{2}$As$_{2}$ points to the
helimagnetic structure proposed above.

The magnetic state of Eu$^{2+}$ moments correlates with the
$\rho(H)$ data shown in Fig. 5. At $H=H_s$=$H^*$ and $T<17$ K, a
turning point can also be found in the $\rho(H)$ curve. This
observation reveals the interplay between SC and the magnetism of
Eu$^{2+}$. For $H>H_s$, the Eu$^{2+}$ spins is fully aligned along
the magnetic field. Thus the magnetic state is basically
homogeneous. Under this circumstance, superconductivity could
survive in the form of superconducting vortices. The electric
current through the sample will result in the dissipative motion of
the vortex, thus showing non-zero resistance. In the HM state
($H<H_s$), one expects non-collinear vortex, which could lead to a
possibly larger dissipation. This is a plausible explanation we can
figure out at present for the resistivity reentrance shown in Fig.
3.

\subsection{Phase Diagram}

Based on the above experimental results, the magnetic and
superconducting phase diagrams were summarized as shown in Fig. 10.
There are five different types of phase regimes. The first is
paramagnetic normal state, located at the upper region in the phase
diagrams. The second is paramagnetic superconducting state, which
has a small area with narrow ranges of temperature and field. In the
third state, located at the lower left side, SC coexists with the
helimagnetic ordering of Eu$^{2+}$ moments. The fourth is FM normal
state, stabilized by external magnetic fields. The last phase shows
the coexistence of SC and FM state, where spontaneous vortex phase
is expected. As can be seen, the phase boundaries are obviously
different for $H\parallel ab$ and $H\parallel c$. However, both
cases show five states in terms of magnetic ordering of Eu$^{2+}$
spins and SC associated with Fe 3$d$ electrons.

\begin{figure}
\includegraphics[width=8cm]{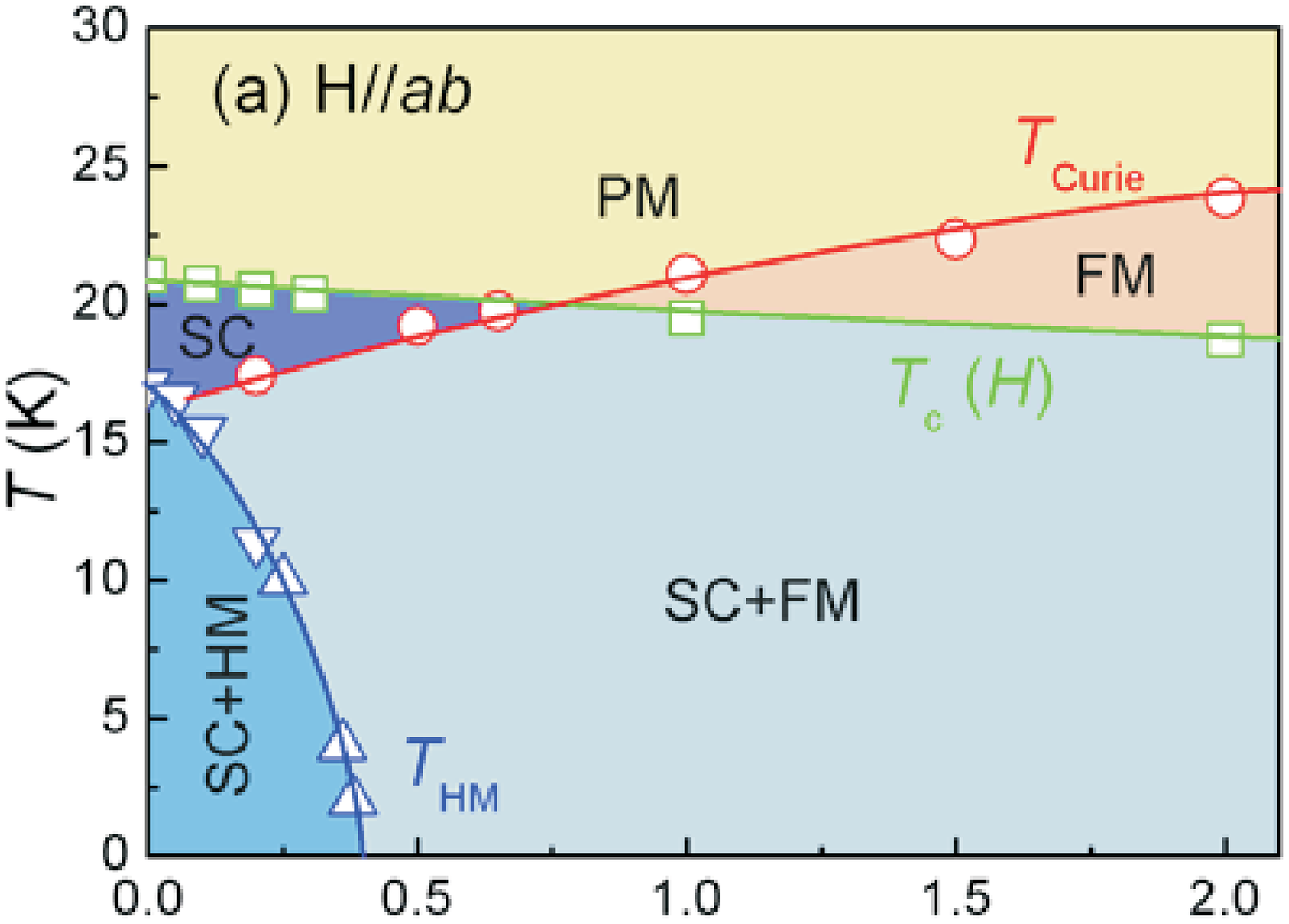}
\includegraphics[width=8cm]{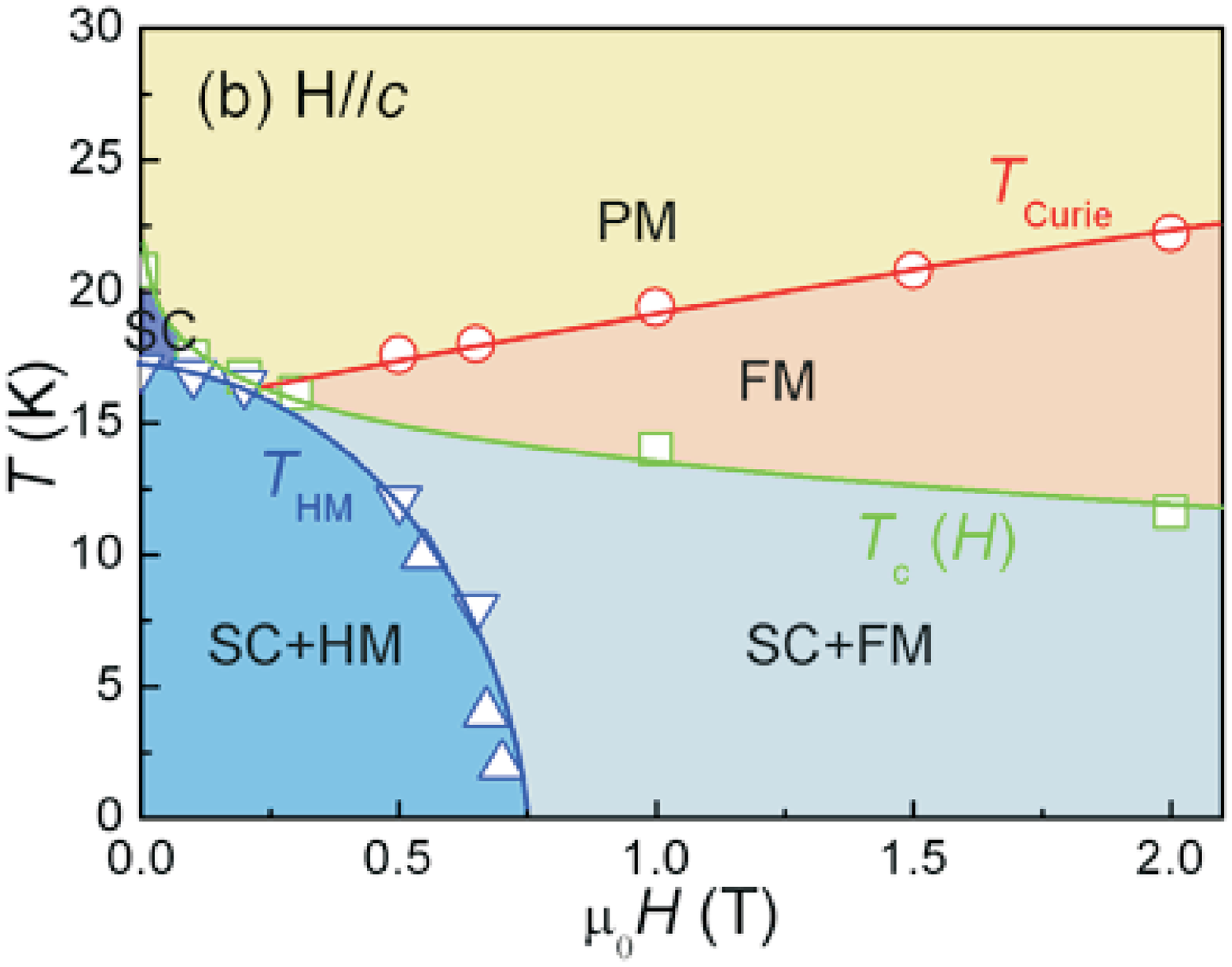}
\caption{(Color online) Electronic phase diagrams in
Eu(Fe$_{0.89}$Co$_{0.11}$)$_{2}$As$_{2}$. PM: paramagnetic state;
SC: superconducting state; FM: ferromagnetic state; SC+HM:
coexistence of superconductivity and helimagnetic order; SC+FM:
coexistence of superconductivity and ferromagnetic state.
(a)$H\parallel ab$; (b)$H\parallel c$.}
\end{figure}

\section{\label{sec:level4}Concluding remarks}

In summary, we have measured the resistivity and magnetization under
magnetic fields on Eu(Fe$_{0.89}$Co$_{0.11}$)$_{2}$As$_{2}$ single
crystals. Evidence of superconducting transition at 21 K was given
from low-field magnetic susceptibility as well as
(magneto)resistivity. Below 17 K, Eu$^{2+}$ moments are most likely
helically ordered under low magnetic fields, which causes
resistivity reentrance. The Eu$^{2+}$ moments can be easily
re-orientated by the external fields, exhibiting the coexistence of
SC and FM state.

There are still some open questions in the present study. One is the
origin of large non-zero resistance. While it is possible that
spontaneous vortex accounts for the non-zero resistance, direct
evidence of spontaneous vortex is called for. The other is the
low-field magnetic susceptibility anomaly at 13 K. Whether it is
truly a PME, and is originated from spontaneous flux is of great
interest. Here we suggest that low-temperature magnetic force
microscopy and scanning SQUID technique should be employed.
Furthermore, specific electrical transport properties such as Hall
coefficient and Nernst coefficient could be helpful to resolve the
above issues.

\begin{acknowledgments}
This work is supported by the NSF of China, National Basic Research
Program of China (No. 2007CB925001) and the PCSIRT of the Ministry
of Education of China (IRT0754).
\end{acknowledgments}

\end{document}